\documentclass[prb,showpacs,preprintnumbers,amsmath,amssymb,showkeys]{revtex4}
\usepackage[pdftex]{color,graphicx}
\usepackage{color,soul}
\definecolor{applegreen}{rgb}{0.29, 0.6, 0.0}

\usepackage{bm}
\usepackage{german}
\usepackage{verbatim}
\usepackage{wasysym}
\usepackage{marvosym}
\usepackage{mathrsfs}
\newcommand{\beq}{\begin{equation}}
	\newcommand{\eeq}{\end{equation}}

\usepackage{wrapfig}
\definecolor{light-gray}{gray}{0.95}
\usepackage{tcolorbox}
\begin{document}

	
\noindent{\LARGE\bf Neues aus dem Kondo-Wunderland}\\
	
\noindent{\Huge\bf Kondomania}\\
	
\noindent{\LARGE S.~KIRCHNER $|$ S.~PASCHEN}\\[2ex]
	
\noindent{\Large\it Vor knapp 90 Jahren wurde der Kondo-Effekt erstmals experimentell beobachtet. Es dauerte 30 Jahre, bis eine theoretische Erkl\"arung gelang. Heute ist Kondo-Physik akueller denn je.}\\[2ex]
	
\noindent{\large Urspr\"unglich bezeichnet der Kondo-Effekt  einen Streumechanismus von Elektronen in Metallen, die St\"orstellen mit internen quantenmechanischen Freiheitsgraden besitzen. Das charakteristische dynamische Wechselspiel  zwischen lokalisierten und itineranten  Zust\"anden tritt in vielen unterschiedlichen Formen auf. Mittlerweile ist klar, dass der Kondo-Effekt die Eigenschaften verschiedenster Klassen von Quantenmaterialien pr\"agt und ein Schl\"ussel f"ur das Verst"andnis ihrer ungew"ohnlichen Eigenschaften ist. Im Folgenden umrei{\ss}en wir eine Auswahl der wichtigsten Entwicklungen dieser Kondomania.}\\[2ex]
		
{\Large\bf Kondo-Effekt}\\

\noindent Einfache Metalle sind vor allem als gute elektrische Leiter bekannt. Das Sommerfeld-Modell beschreibt ihre Leitungselektronen als ein wechselwirkungsfreies Gas, in dem die Elektronen auf beachtlich hohe Geschwindigkeiten kommen. Das ist durch die Quantisierung ihrer Zust\"ande und das Pauliverbot, wonach ein Zustand maximal mit zwei Elektronen besetzt sein kann, bedingt. Das Pauliverbot f\"uhrt zu einem Grundzustand, der durch ein sukkzessives Auff\"ullen der erlaubten Impulseigenzust\"ande gegeben ist. Die besetzten Zust\"ande mit der h\"ochsten Energie im Grundzustand, der Fermi-Energie, bilden eine Fl\"ache im Impulsraum, die Fermi-Fl\"ache genannt wird. Gitterfehlstellen, Fremdatome oder auch kollektive Anregungen k"onnen zur Streuung der Elektronen f"uhren \cite{Ros99.1}. 

\begin{figure}
	\includegraphics[width=.8\textwidth]{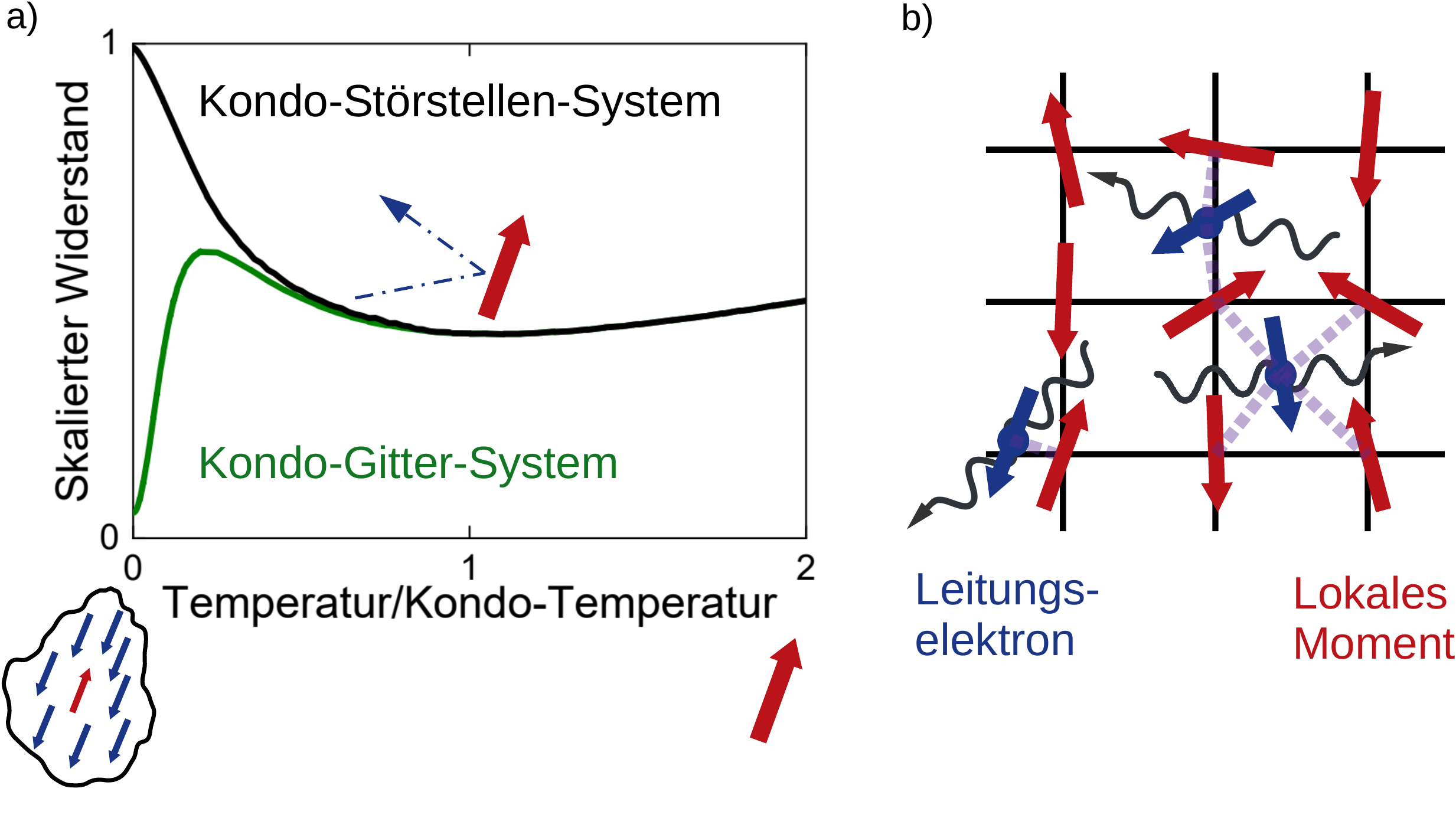}
	\caption{\textbf{KONDO-WIDERSTAND UND KONDO-GITTER-MODELL\\
	a) Charakteristische Temperaturabh\"angigkeit des elektrischen Widerstands im Kondo-St\"orstellen- (schwarz) und Kondo-Gitter-Modell (gr\"un). F\"ur Temperaturen weit oberhalb der Kondo-Temperatur $T_{\rm K}$ ist der Spin im wesentlichen frei, w\"ahrend er f\"ur $T\rightarrow 0$ durch die Leitungselektronen abgeschirmt ist (Symbole unten). b) Schema eines einfachen Kondo-Gitter-Modells. An jedem Gitterplatz findet Kondo-Streuung (strichlierte Linien) zwischen lokalen Momenten (rot) und den Spins der Leitungselektronen (blau) statt.}}
	\label{Fig_Kondo}
\end{figure}

Es ist charakteristisch f\"ur metallische Systeme, dass die Streurate mit steigender Temperatur zunimmt. Falls ein Metall jedoch St\"orstellen mit intrinsische Freiheitsgraden besitzt, die an die Leitungselektronen ankoppeln, ist dies anders: die Streurate und damit auch der elektrische Widerstand steigt mit {\em fallender} Temperatur (Abbildung \ref{Fig_Kondo}\,a). Im einfachsten Fall ist die St\"orstelle ein Spin, der \"uber die Austauschwechselwirkung an die Spins der Leitungselekronen ankoppelt. Andere Arten von St\"orstellen werden wir sp\"ater noch kennenlernen. Die Ursache f\"ur diesen Widerstandanstieg bei tiefen Temperaturen ist der Kondo-Effekt. Er ist nach dem japanischen Physiker Jun Kondo benannt, der in den 1960er Jahren zeigte, dass Vielteilchenprozesse f\"ur dieses Verhalten verantwortlich sind \cite{Hewson}.

Vielteilchenprozesse im Metall betreffen Leitungselektronen in der N\"ahe der Fermi-Energie; man bezeichnet sie als Teilchen-Loch-Anregungen. Im Falle der Kondo-Wechselwirkung versuchen sie, den lokalen Freiheitsgrad abzuschirmen, indem sie ein Vielteilchen-Singulett mit ihm bilden (linker Inset Abbildung \ref{Fig_Kondo}\,a). Die Bindungsenergie dieses Singuletts ist die Kondo-Energie, die entsprechende Temperatur die Kondo-Temperatur ($T_{\rm K}$). Sie ist die einzige relevante Energieskala des Problems. Widerstandskurven als Funktion des Verh\"altnisses aus Temperatur und Kondo-Temperatur aufgetragen (wie in Abbildung \ref{Fig_Kondo}\,a) sind daher universell. F\"ur Temperaturen weit unterhalb der Kondo-Temperatur findet nur Potentialstreuung statt, wie man es f\"ur Metalle mit statischen Streuzentren erwartet. Tats\"achlich erreicht der f\"ur $T\rightarrow 0$ verbleibende Potentialstreuterm den maximal m\"oglichen Wert (unit\"arer Limes).
		
Der Kondo-Effekt ist f\"ur die ungew\"ohnlichen Eigenschaften einer ganzen Materialklasse, den Schwerfermionen-Systemen verantwortlich. Er spielt zudem f\"ur theoretische Entwicklungen eine bedeutende Rolle. So stellte seine Beschreibung die erste erfolgreiche Anwendung von Renormierungsgruppen-Rechnungen in der Festk\"orperphysik dar (Infobox\,1). Weiterhin hat sich herausgestellt, dass der Kondo-Effekt ein Vielteilcheneffekt ist, der mit einer Vielzahl von theoretischen Methoden behandeln l\"asst. Dazu geh\"ohren exakte Behandlungen, die allerdings nur in einem engen Rahmen G\"ultigkeit besizten, Methoden der Quantenfeldtheorie, sowie semi-analytische und rein numerische Methoden. Ausserdem f\"uhren Standardmethoden zur Beschreibung stark korrelierter Elektronensysteme, wie z.B.\ die dynamische Molekularfeld-Theorie, kurz DMFT,  oft auf Modelle, die den Kondo-Effekt zeigen. Schlie{\ss}lich kann (wie weiter unten gezeigt) der urspr\"ungliche Spin-Kondo-Effekt auf eine Reihe von Situationen verallgemeinert werden, die scheinbar wenig mit magnetischen Momenten in Metallen gemein haben, womit sein G\"ultigkeitsbereich nochmal betr\"achtlich w\"achst.\\[2ex]

{\Large\bf Kondo-Gitter}\\

\noindent Das f\"ur den Kondo-Effekt charakteristische Wechselspiel von lokalisierten und delokalierten Freiheitsgraden tritt auch auf, wenn eine periodische Anordnung von lokalisierten Momenten an ein Band von Leitungselektronen ankoppelt. Das ist in Schwerfermionen-Systemen der Fall. Dies sind intermetallischen Verbindungen, die neben $s$, $p$ und $d$-Elementen auch Lanthanoide ($4f$-Elemente) wie Cer, Samarium oder Ytterbium oder Actinoide ($5f$-Elemente) wie Uran oder Plutonium enthalten. Die Elektronen in diesen nur teilweise gef\"ullten, energetisch tieferliegenden Orbitalen nehmen nicht direkt an der metallischen Bindung teil und bilden lokalisierte Momente aus. Sind dies keine reinen Spin-Momente, m\"ussen weitere Energieskalen wie die Spin-Bahn-Kopplung und Kristallfeldaufspaltungen der orbitalen Entartung ber\"ucksichtigt werden. Solche Schwerfermionen-Systeme werden mit dem Kondo-Gitter-Modell (Abbildung \ref{Fig_Kondo}\,b) beschrieben.

\begin{figure}[!htb]
	\centering
	\includegraphics[width=0.7\linewidth]{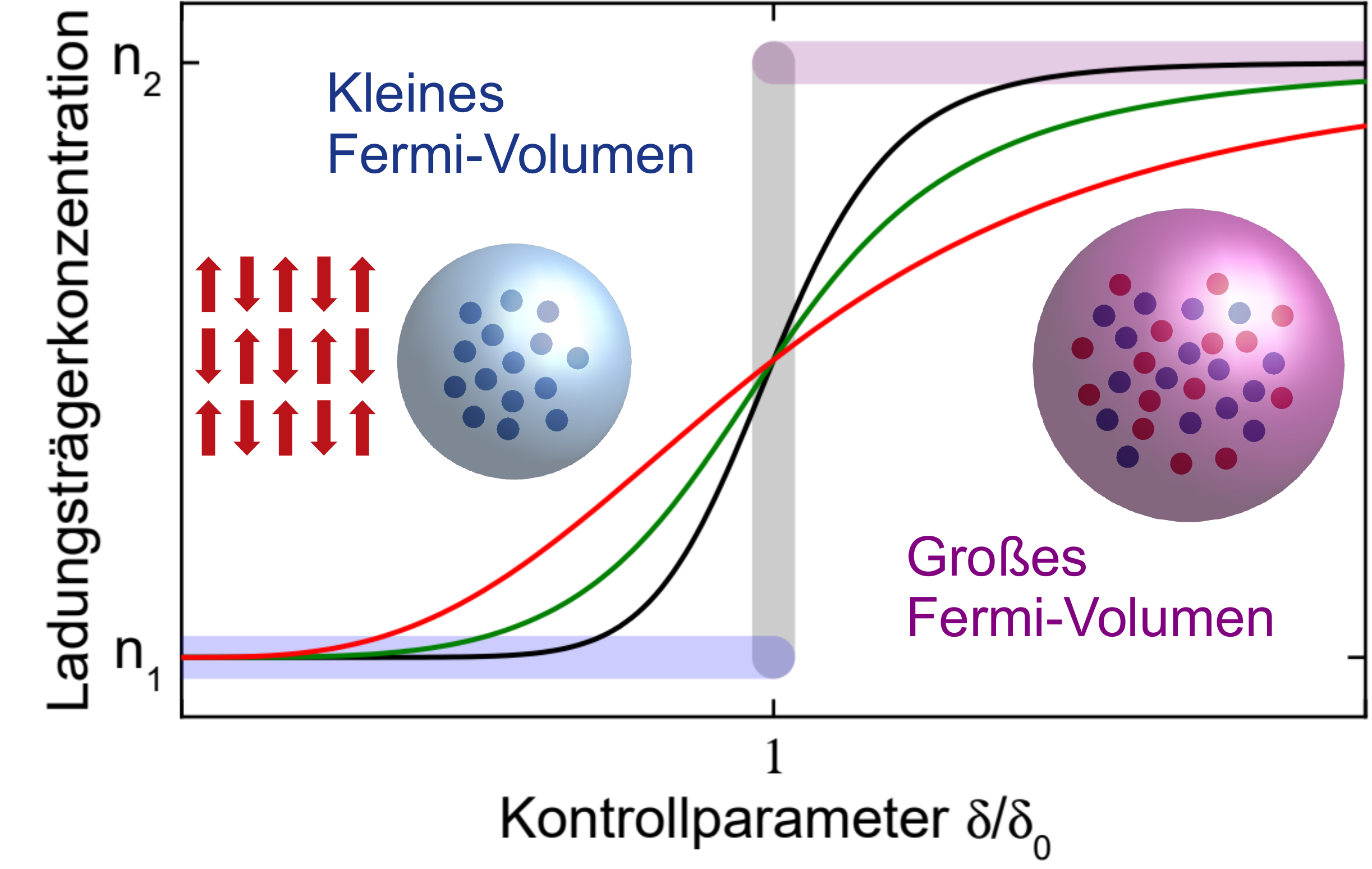}
	\caption{\bf{HALLEFFEKT BEIM KONDO-AUFBRUCH\\
	Die aus Messungen des Halleffekts bestimmte Ladungstr\"agerkonzentration \"andert sich als Funktion eines Kontrollparameters $\delta$ an einem quantenkritischen Punkt ($\delta = \delta_{\rm c}$) mit Kondo-Aufbruch sprungartig von einem kleineren Wert $n_1$, der nur die Leitungselektronen (blau) umfasst, auf einen gr\"o{\ss}eren Wert $n_2$, zu dem auch die zuvor lokalisierten Elektronen (rot) beitragen (dicke Linien). Messkurven (schmale Linien) zeigen mit abnehmender Temperatur (von rot nach schwarz) einen immer sch\"arferen \"Ubergang. $\delta_0$ ist der Mittelpunkt dieses \"Ubergangs bei endlicher Temperatur.}}
	\label{Fig_Hall}
\end{figure}

Hier sitzt auf jedem Gitterplatz ein lokalisiertes Moment, das wie im St\"orstellenfall mit den Spins der Leitungselektronen \"uber die Austauschkopplung wechselwirkt. In einem der m\"oglichen Grundzust\"ande, in den das System mit abnehmender Temperatur \"ubergehen kann, ist jedes lokalisierte Moment durch die Spins der Leitungselektronen v\"ollig abgeschirmt. Die bei tiefen Temperaturen resultierenden elektronischen Anregungen entsprechen in diesem Fall schweren Quasiteilchen (ihre Masse kann mehr als die tausendfache Masse freier Elektronen erreichen), woraus auch der Name dieser Materialklasse resultiert \cite{Eckern.95}. Jedes lokale Moment bildet in diesem Fall eine Kondo-Resonanz an der Fermi-Energie aus und nimmt \"uber diese Resonanz an den elektronischen Eigenschaften des Gesamtsystems teil. Das Fermi-Volumen (Volumen, das von der Fermi-Fl\"ache eingeschlossen wird) vergr\"o{\ss}ert sich (rechter Inset in Abbildung \ref{Fig_Hall}). Die bei hohen Temperaturen lokalisierten Momente leisten einen makroskopisch gro{\ss}en Beitrag zur Gesamtentropie. Dieser muss beim Abk\"uhlen ($T\rightarrow 0$) freigesetzt werden, was sich in der W\"armekapazit\"at niederschl\"agt. Da in diesem Zustand die elektronische Zustandsdichte am Fermi-Niveau stark erh\"oht ist, kommt es in vielen Schwerfermionen-Verbingungen zu (itineranter) magnetischer Ordnung, \"ahnlich wie dies von $d$-Elementen wie metallischem Eisen bekannt ist. 

Allerdings ist dies nicht die einzige Art, wie in einem Kondo-Gitter magnetische Ordnung entstehen kann. Unter bestimmten Umst\"anden kann es auch zu magnetischer Ordnung von lokalisierten Momenten kommen. Um dies zu verstehen, m\"ussen wir einen weitere Wechselwirkung einf\"uhren, die Ruderman-Kittel-Kasuya-Yosida oder RKKY-Wechselwirkung. Sie ist eine indirekte Austauschkopplung zwischen den lokalen Momenten, die auf der Austauschkopplung mit den Leitungselektronen basiert, und favorisiert magnetische Ordnung. Ob ein System magnetisch ordnet oder nicht, h\"angt von der relativen Gr\"o{\ss}e der Kondo- und RKKY-Wechselwirkung ab. Dominiert die RKKY-Wechselwirkung, tritt Ordnung auf (es sei denn geometrische Frustration oder Unordnung verhindern dies). Der metallische Grundzustand beinhaltet dabei nur die reinen Leitungselektronen und das Fermi-Volumen ist ``klein'' (linker Inset in Abbildung \ref{Fig_Hall}). Kondo-Streuprozesse sind in diesem Fall bei endlichen Temperaturen zwar vorhanden (und k\"onnen auch die effektive Masse der Quasiteilchen erh\"ohen), sie f\"uhren aber nicht zur vollst\"andigen Abschirmung der lokalen Freiheitsgrade.\\[2ex]

{\Large\bf Kondo im Aufbruch}\\
		
\noindent Der Kondo-Effekt bildet sich mit abnehmender Temperatur kontinuierlich aus. Anders als bei magnetischer Ordnung oder Supraleitung gibt es also keine Kondo-Phase und auch keinen (thermischen) Kondo-Phasen\"ubergang. Umso interessanter ist es, dass sich dies beim absoluten Temperaturnullpunkt anders verh\"alt. Eine Vielzahl von Experimenten weist darauf hin, dass der Kondo-Effekt an einem Quantenphasen\"ubergang instantan aufbrechen kann \cite{Kir20.1,Pas21.1}. Besonders anschaulich zeigt dies der Halleffekt. Er gibt Auskunft \"uber die Anzahl beweglicher Ladungstr\"ager bzw.\ \"uber das Fermi-Volumen. Ist der Kondo-Effekt intakt, so sind sowohl die (urspr\"unglich) lokalisierten magnetischen Momente als auch die Leitungselektronen im Fermi-Volumen enthalten. Bricht er auf, so bleiben nur die Leitungselektronen zur\"uck. In Experimenten, die nat\"urlich nur bei endlichen Temperaturen durchgef\"uhrt werden k\"onnen, wurde dies als eine \"Anderung des Hallkoeffizienten als Funktion eines nichtthermischen Kontrollparameters (z.B.\ des Magnetfeldes \cite{Pas21.1}) detektiert, wobei diese \"Anderung mit abnehmender Temperatur immer abrupter passiert und im Grenzfall von $T=0$ instantan (Abbildung \ref{Fig_Hall}).
		
\begin{figure}[!htb]
	\centering
	\includegraphics[width=0.35\linewidth]{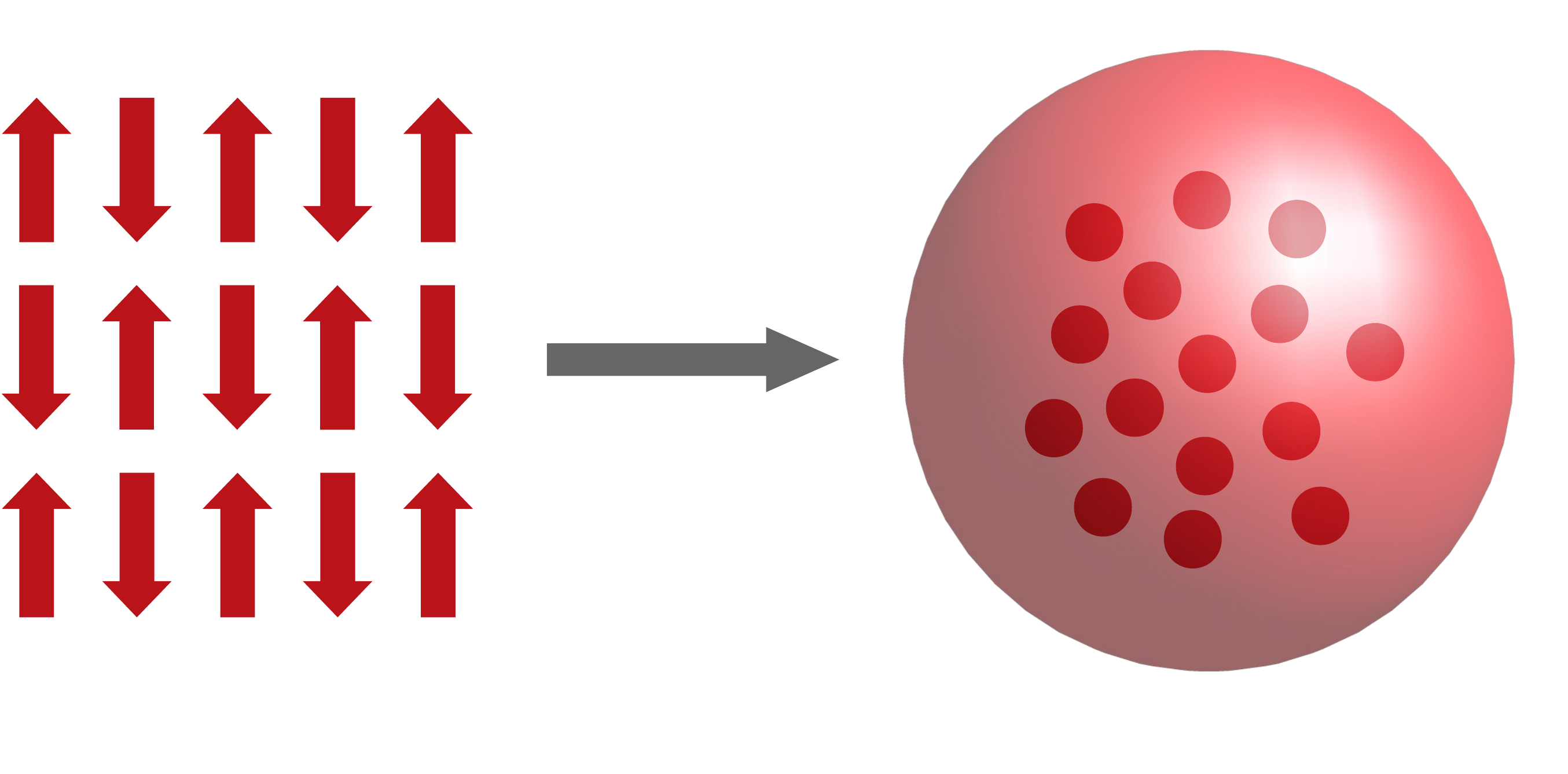}
	\caption{\bf{(MOTT) METALL-ISOLATOR-\"UBERGANG}}
	\label{Fig_Mott}
\end{figure}

Da in einem Kondo-Metall auf beiden Seiten eines solchen Quantenphasen\"ubergangs Leitungselektronen vorliegen, ist dies ein Metall--Metall-\"Ubergang. Falls hingegen alle Leitungselektronen beim \"Ubergang lokalisieren, tritt ein (Mott) Metall--Isolator-\"Ubergang auf, wie er zum Beispiel von \"Ubergangsmetalloxiden her bekannt ist (Abbildung \ref{Fig_Mott}).
		
Genauere Untersuchungen zeigen, dass diese Elektronenlokalisierung bzw.\ Delokalisierung kein statischer, sondern ein h\"ochst dynamischer Prozess ist, dem quantenkritische Fluktuationen zugrunde liegen. Am quantenkritischen Punkt liegt das Material quasi in zwei unterschiedlichen Zust\"anden gleichzeitig vor (eben zum Beispiel einem mit gro{\ss}em und einem mit kleinem Fermi-Volumen) und fluktuiert zwischen beiden hin und her. Zur Untersuchung dieser Dynamik eignen sich Messungen, die als Funktion der Energie bzw.\ der Frequenz durchgef\"uhrt werden k\"onnen. Beispiele sind die optische Leitf\"ahigkeit oder die inelastische Neutronenstreuung. Bei verschiedenen Temperaturen und Anregungsenergien gemessene Daten zeigen je nach der Art des quantenkritischen Punktes bestimmte Skalenverhalten [Infobox Quantenkritikalit\"at und dynamisches Skalenverhalten].
		
\begin{figure}[!h]
    \centering
    \includegraphics[width=0.5\linewidth]{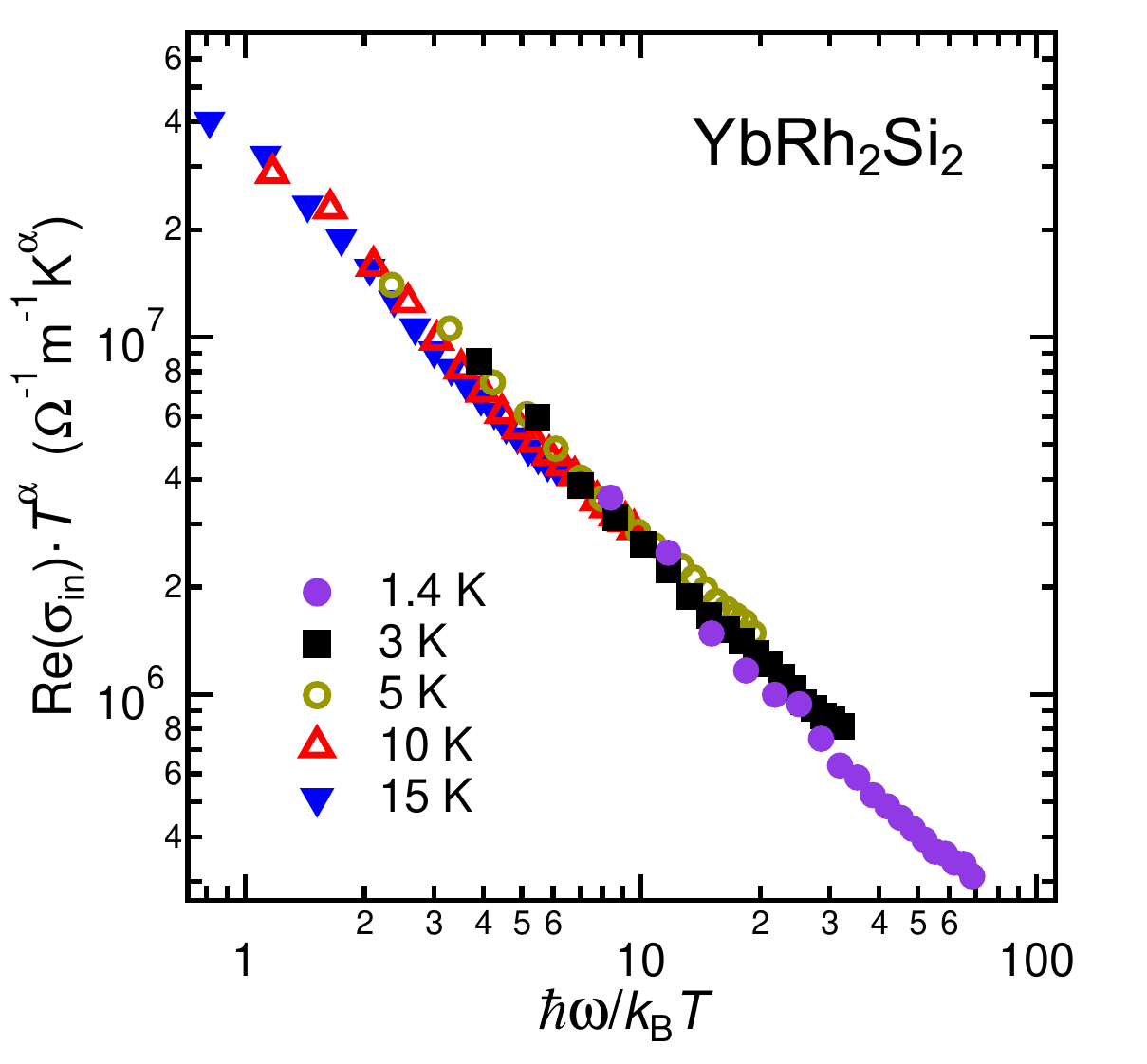}
    \caption{{\bf OPTISCHE LEITF\"AHIGKEIT BEIM KONDO-AUFBRUCH\\
	Wird der Realteil der (inelastischen) Terahertz-Leitf\"ahigkeit $\sigma_{\rm in}$ von YbRh$_2$Si$_2$ wie gezeigt aufgetragen, so fallen alle Daten auf eine universelle Kurve. $\hbar\omega$ ist die Energie der Terahertz-Strahlung, $k_{\rm B}T$ die thermische Energie. Der Temperaturexponent $\alpha$ ist 1} (nach \cite{Pro20.1}).}
	\label{Fig_opt}
\end{figure}

Universelles Verhalten als Funktion des Verh\"altnisses aus Energie und Temperatur, wie es f\"ur die Schwerfermionen-Verbindung YbRh$_2$Si$_2$ auftritt (Abbildung \ref{Fig_opt}), ist ungew\"ohnlich. F\"ur Quntenphasen\"uberg\"ange, deren Verhalten zur G\"anze durch das Verschwinden eines Ordnungsparameters (im Falle von Magneten der Magnetisierung) bestimmt ist, ist dieses Verhalten gem\"a{\ss} theoretischen Betrachtungen nicht erlaubt. Tats\"achlich geht in YbRh$_2$Si$_2$ der Wechsel von einem antiferromagnetischen in den paramagnetischen Zustand mit der oben beschriebenen Delokalisierung von Elektronen einher. Das Auftreten der ``Energie-\"uber-Temperatur-Skalierung'' bedeutet, dass die Ladungsdelokalisierung Teil des quantenkritischen Verhaltens ist. Dieses ist also neben den Fluktuationen des Ordnungsparameters durch die Fluktuationen der Ladungstr\"agerdelokalisierung bedingt\cite{Pro20.1,Kir20.1,Pas21.1}.
		
Die genaue Form des Skalenverhaltens der optischen Leitf\"ahigkeit, in diesem Fall mit einem kritischen Exponenten $\alpha = 1$, entspricht einem in der Temperatur linearen elektrischen Widerstandsverlauf, wie er tats\"achlich in YbRh$_2$Si$_2$ auftritt. F\"ur ``normale'' Metalle wird selbst im Fall von starken elektronischen Korrelationen zumeist ein quadratisches Temperaturverhalten beobachtet, das im Einklang mit der Theorie der Fermifl\"ussigkeit ist \cite{Ros99.1}. Abweichungen davon werden als Nicht-Fermi-Fl\"ussigkeitsverhalten bezeichnet. Einem linearen Verlauf kommt dabei besondere Bedeutung zu. Er wird in ganz unterschiedlichen Materialklassen mit starken Elektronenkorrelationen beobachtet, und zwar oft im Zusammenhang mit unkonventioneller Supraleitung.\\[2ex]
\newpage

{\Large\bf Kondo-Effekt und Supraleitung}\\
		
\noindent Das Ph\"anomen der Supraleitung -- gekennzeichnet durch das Verschwinden des elektrischen Widerstands und perfekten Diamagnetismus -- fasziniert Experten und Laien gleichermassen. In der einfachsten Form von Supraleitung werden zwei Elektronen \"uber den Austauch einer Gitterschwingung (eines Phonons) zu einem supraleitenden Cooper-Paar zusammengehalten. Dies wird als konventionelle Supraleitung bezeichnet. In unkonventionellen Supraleitern erfolgt die Cooper-Paarbildung \"uber andere Austauschteilchen. Vergleichsweise gut verstanden ist Supraleitung durch Spin-fluktuationen, wie sie beim Verschwinden eines magnetischen Ordnungsparameters in der N\"ahe eines (konventionellen) quantenkritischen Punktes auftreten \cite{Ros99.1}. Allerdings scheint die interessanteste Art der Supraleitung, wie sie auch in den Kuprat-Hochtemperatursupraleitern vorkommt, nicht mit diesem Mechanismus kompatibel: einerseits, da die Supraleitung hier nicht am Rand einer magnetisch geordneten Phase zu entstehen scheint und andererseits, da das lineare Termperaturverhalten nicht zu den theoretische Vorhersagen dieses Modells passt.
		
\begin{figure}[!h]
    \centering
	\includegraphics[width=0.5\linewidth]{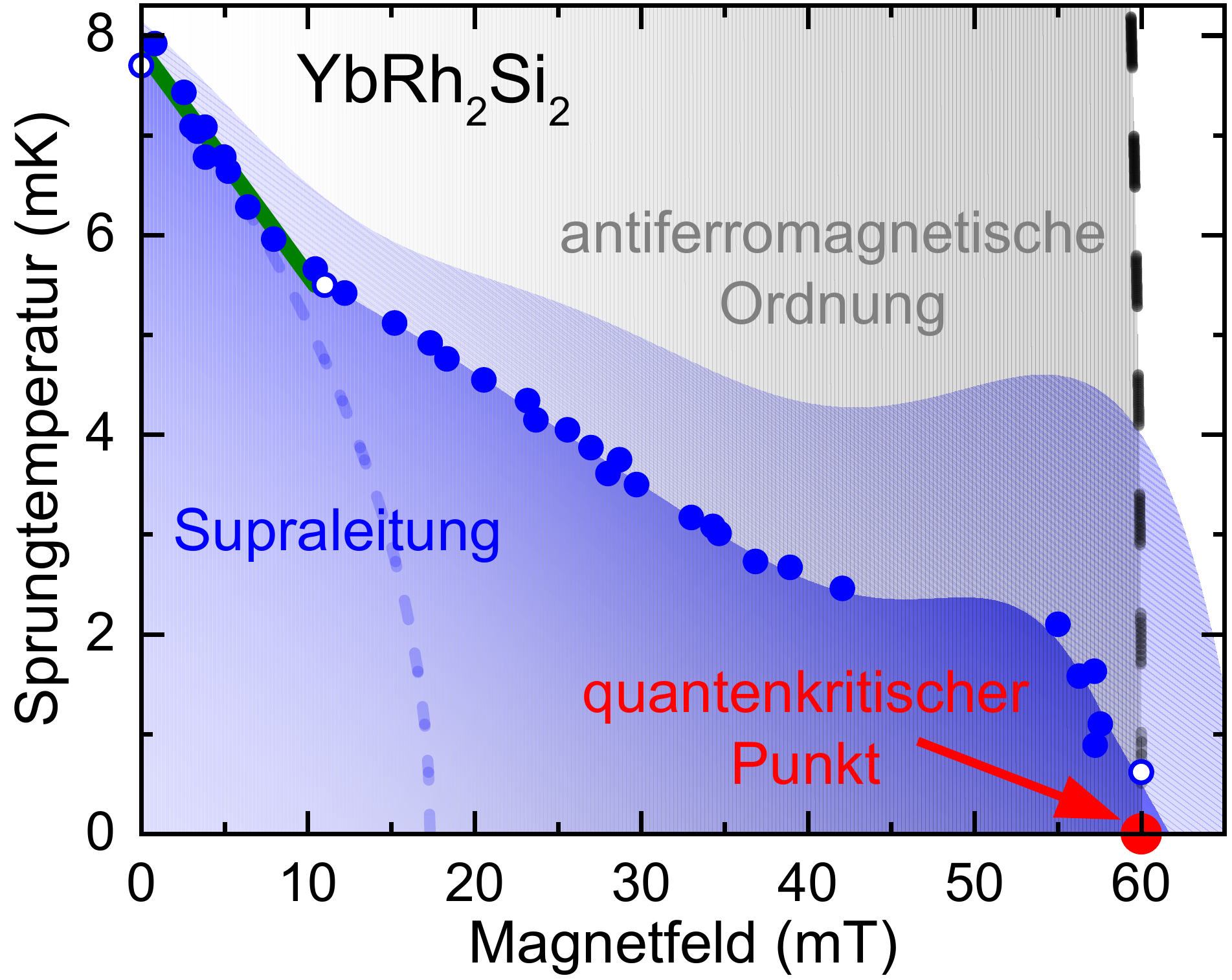}
	\caption{{\bf SUPRALEITUNG BEIM KONDO-AUFBRUCH\\
	Die supraleitende Sprungtemperatur wurde mittels elektrischen Widerstandsmessungen als jene Temperatur bestimmt, bei der der Widerstand auf 50\% Prozent seines Wertes im normalleitenden Zustand direkt oberhalb des \"Ubergangs gefallen ist (blauen Punkte). Als Grenze des blau schattierten Bereichs ist ausserdem die Temperatur gezeigt, bei der der Widerstand auf 90\% des normalleitenden Wertes gefallen ist. Als Magnetfeld wird hier das Feld, multipliziert mit der Permeabilit\"at des Vakuums $\mu_0$ bezeichnet (die magnetische Flussdichte). Am quantenkritischen Punkt zeigt der Widerstand oberhalb der Supraleitung lineares Verhalten \"uber mehr als drei Gr\"o{\ss}enordnungen in der Temperatur} (nach \cite{Ngu21.1}).}
	\label{YRS_sc}
\end{figure}

Der Fage, ob die beim Kondo-Aufbruch erzeugten quantenkritischen Fluktuationen zu Supraleitung f\"uhren k\"onnen, wurde in einer k\"urzlich ver\"offentlichten Studie im wahrsten Sinne des Wortes auf den Grund gegangen. Hier wurde der elektrische Widerstand von YbRh$_2$Si$_2$ bei Temperaturen bis unter ein Millikelvin gemessen und tats\"achlich Supraleitung gefunden, die unmittelbar aus dem linearen Temperaturverlauf bei h\"oheren Temperaturen entsteht (Abbildung \ref{YRS_sc}). Dies legt die Vermutung nahe, dass die Supraleitung durch dieselben Quantenfluktuationen bedingt ist wie der lineare Widerstand.

Interessant ist auch die Form der supaleitenden Phase. Nach einem erst rascheren und dann graduellen Abfall der Sprungtemperatur mit steigendem Magnetfeld kommt es in der N\"ahe des quantenkritischen Punktes zu einem Plateau bzw.\ sogar zu einem Wiederanstieg (vgl.\ 90\%-Kurve in Abbildung \ref{YRS_sc}). Dies k\"onnte bedeuten, dass hier zwei supraleitende Phasen mit unterschiedlichen Typen von Cooper-Paaren vorliegen: eine mit der \"ublichen antiparallelen Spineinstellung (``Spin-Singulett'') und eine mit der gegen Magnetfelder wesentlich robusteren parallelen Spineinstellung (``Spin-Triplet''), wobei letzterer Zustand von gro{\ss}em Interesse f\"ur bestimmte Quantentechnologien ist (siehe weiter unten).

In den Kuprat-Hochtemperatursupraleitern sind die f\"ur die Supraleitung verantwortlichen Kupferoxid-Ebenen im reinen (st\"ochiometrischen) Zustand Mott-Isolatoren. Das Dotieren mit L\"ochern f\"uhrt zu Zhang-Rice-Singuletts, die f\"ur den Ladungstransport im dotierenten Isolator verantwortlich sind. Eine Anzahl von Experimenten weist darauf hin, dass in der N\"ahe der optimalen Supraleitung ein Delokalisierungs\"ubergang auftritt, \"ahnlich der oben beschriebenen Delokalisierung am Kondo-Aufbruch-Quantenphasen\"ubergang. Ob damit auch ein \"ahnlicher Mechanismus der Cooper-Paarung einhergeht, ist Gegenstand der aktuellen Forschung.\\[2ex]

		
{\Large\bf  Kondo-Effekt ganz ohne Spin}\\
	
\noindent Bisher war unsere Diskussion des internen Freiheitsgrads an den Spin und seine mit der Zeitumkehrsymmetrie verbundene Entartung gekn\"upft. Wird diese Entartung (merklich) aufgehoben, kann sich der Kondo-Effekt nicht ausbilden. Dies ist zum Beispiel in einem hinreichend gro{\ss}en Magnetfeld der Fall ($g\mu_{\rm B}B\gg k_{\rm B}T_{\rm K}$, wobei $B$ die magnetische Induktion, $g$ der Land\'e-Faktor, $\mu_{\rm B}$ das Bohrsche Magneton und $k_{\rm B}$ die Boltzmann-Konstante ist). 
Die besondere Rolle von Symmetrien im Kondo-Effekt l\"a{\ss}t sich am Beispiel des Doppelmuldenpotentials verdeutlichen. 

\begin{figure}[!h]
	\centering
	\includegraphics[width=0.8\linewidth]{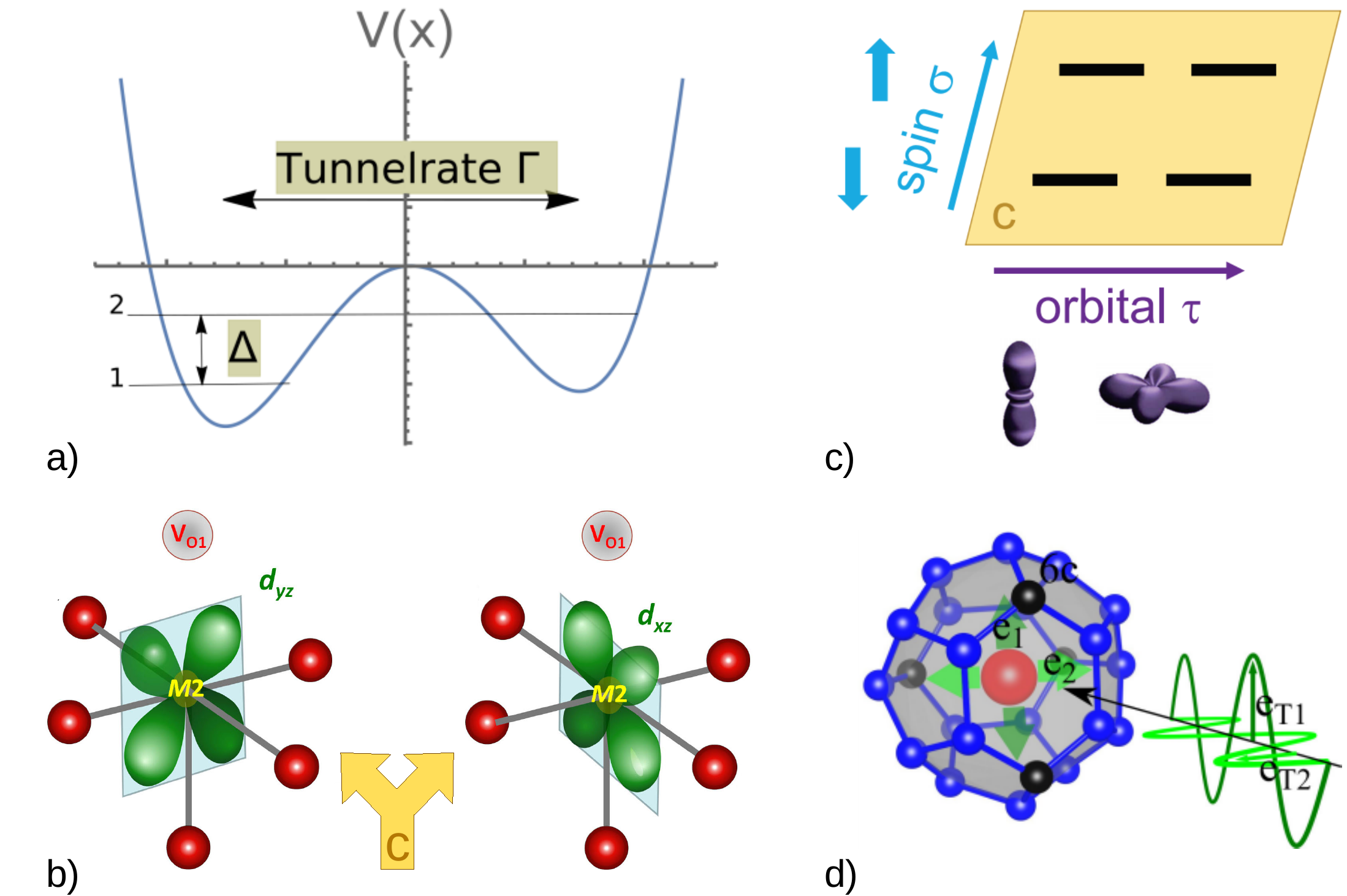}
	\caption{{\bf VERALLGEMEINERTER KONDO-EFFEKT\\
	a) Die beiden tiefstliegenden Zust\"ande des Doppelmuldenpotentials lassen sich durch die im linken (``$\downarrow$'') und rechten (``$\uparrow$'') Minimum lokalisierten Zust\"ande darstellen. \"Uberg\"ange zwischen $\uparrow$ und $\downarrow$ werden durch $\Gamma S^x$  mit Hilfe des Pseudospinoperators $S^x$ beschrieben. Dies bildet die Basis f\"ur eine Pseudospinalgebra. b) Der orbital Kondo-Effekt in RuO$_2$ entsteht durch Sauerstofffehlstellen (V$_{\rm O2}$) in der Rutilstruktur und basiert auf der Entartung der $4d_{xz}$ und $4d_{yz}$-Orbitale der benachbarten
	Ru-Ionen} (nach \cite{Yeh2020}). {\bf c) In Kristallen mit hoher Symmetrie kann die $4f$-Wellenfunktion einen 4-fach entarteten Grundzustand haben. Dann bilden der Spin und der orbitale Pseudospin gemeinsam den lokalisierten Freiheitsgrad} (nach \cite{Pas21.1}). {\bf d) Lokale zweifach entartete Schwingungsmoden wie sie in bestimmten K\"afigstrukturen auftreten, k\"onnen ebenfalls einen Pseudospin definieren. In diesem Fall k\"onnen akustische Phononen, die eine lineare Dispersion besitzen, als effektives Bad dienen} (nach \cite{Ike19.1}).}
	\label{Fig_other_Kondo}
\end{figure}

Dieses Potential wird oft verwendet, um das Tunneln von Defektatomen zwischen zwei Positionen zu beschreiben (Abbildung \ref{Fig_other_Kondo}\,a). Zur  Modellierung dieser Situation kann ein Pseudospinoperator eingef\"uhrt werden, der statt der beiden Spin-1/2-Zust\"ande im Spin-Kondo-Effekt die beiden m\"oglichen Atompositionen beschreibt. Das Hin- und Hertunneln wird dann durch einen weiteren Pseudospinoperator ($S^x$) vermittelt und die Energieaufspaltung zwischen den beiden Energieniveaus durch einen effektiven Zeeman-Term ($\sim \Delta S^z$) beschrieben. Die Streuung von Leitungselektronen an einem solchen Zwei-Niveau-System kann  \"Uberg\"ange zwischen den beiden lokalen Zust"anden induzieren, was in der Darstellung des Pseudospins einem Austauschprozess wie im Kondo-Modell entspricht. Es zeigt sich, dass f\"ur Tunnelsysteme, in denen die Entartung der lokalen Zust\"ande nicht durch eine Symmetrie gesch"utzt ist, der dynamisch erzeugte effektive Zeeman-Term viel gr\"o{\ss}er als die Kondo-Energie ist, sodass kein Kondo-Effekt auftreten kann\cite{2CKReview}. 

\begin{figure}[!h]
	\centering
	\includegraphics[width=0.5\linewidth]{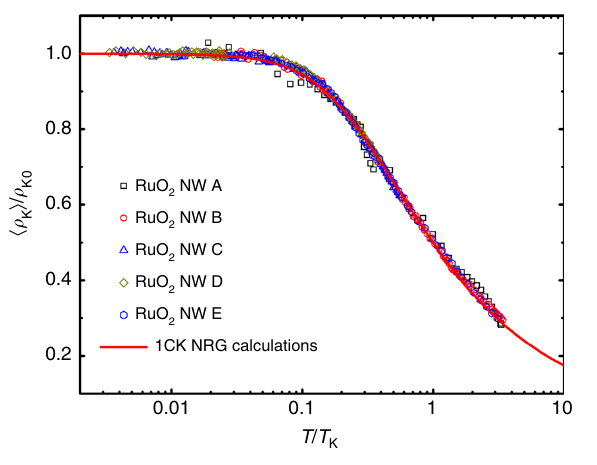}
	\caption{{\bf ORBITALER KONDO EFFEKT IN RuO$_2$\\
	Der auf seinen Wert bei den tiefsten Temperaturen normierte elektrische Widerstand von RuO$_2$-Dr\"ahten, in denen Sauerstoffst\"orstellen einen orbitalen Kondo-Effekt induzieren, zeigt universelles Verhalten, wobei die Kondo-Temperatur $T_{\rm K}$ die einzig relevante Energieskala ist} (nach \cite{Yeh2020}).}
	\label{Fig_ruo2}
\end{figure}

In Materialien, deren Kristallstrukturen eine entsprechend hohe Punktsymmetrie aufweisen, tritt hingegen kein solcher Zeeman-Term auf und der Kondo-Effekt kann sich ausbilden. Ein Beispiel hierf\"ur ist RuO$_2$, wo aufgrund der Entartung der beiden 4$d$-Orbitale ein orbitaler Kondo-Effekt auftritt (Abbildung \ref{Fig_other_Kondo}\,b). Abbildung \ref{Fig_ruo2} zeigt den Widerstand von RuO$_2$ aufgrund des orbitalen Kondo-Effekts und einen Vergleich mit Rechnungen, die auf dem Kondo-Modell beruhen.

Da Leitungselektronen Spin-1/2-Teilchen sind und ohne \"au{\ss}eres Magnetfeld oder magnetische Ordnung Spin-Entartung vorliegt (ein hypothetisches Spin-``up''- und Spin-``down''-Leitungsband haben die gleiche Energie), erwartet man eine Besonderheit f\"ur den orbitalen Kondo-Effekt: Die beiden Spinkomponenten der Leitungselektronen versuchen unabh\"angig voneinander ein Pseudospin-Singulett mit dem lokalen Freiheitsgrad zu bilden. Es kommt zu einem Frustrationseffekt und zur \"Uberabschirmung. Als Folge davon erwartet man die Ausbildung des sogenannten Zweikanal-Kondo-Zustands \cite{2CKReview}. Tats\"achlich wurde dieser in dem zu RuO$_2$ analogen aber  nichtmagnetischen Material IrO$_2$ \"uber eine wurzelf\"ormige Temperaturabh\"angigkeit des elektrischen Widerstands bei tiefen Temperaturen nachgewiesen. RuO$_2$ hingegen ordnet magnetisch und der Zweikanal-Kondo-Effekt tritt nicht auf. In beiden Materialien hat ein angelegtes Magnetfeld keinen Einfluss auf die Kondo-Streuung, was die nichtmagnetische Natur des hier beobachteten Effekts weiter belegt \cite{Yeh2020}.

Es sind auch Materialien bekannt, in denen sich sowohl der magnetische als auch der orbitale Kondo-Effekt ausbilden. Hierzu muss das lokale Multiplett eine hinreichend hohe Entartung aufweisen (was zum Beispiel f\"ur das $4f^1$-Elektron von Cerium in lokaler Punktsymmetrie der Fall sein kann) und  durch Spin und orbitale Quantenzahlen charakterisiert sein (Abbildung \ref{Fig_other_Kondo}\,c).

Eine weitere nichtmagnetische Version des Kondo-Effekts ist der Ladungs-Kondo-Effekt. Er kann entstehen, wenn Entartung zwischen einem leeren und einem doppelt besetzen lokalen Zustand vorliegt. Damit diese beiden Zust\"ande das Grundzustands-Dublett bilden, muss die effektive lokale Coulomb-Absto{\ss}ung negativ sein. Das kann z.B.\ bei Ankopplung an ein lokales Phonon der Fall sein. Ungew\"ohnliche Transporteigenschaften in mit Thallium dotiertem Bleitellurid (PbTe) wurden als Evidenz f\"ur diese Physik interpretiert. Au{\ss}erdem k\"onnte der Ladungs-Kondo-Effekt auch f\"ur die in diesem System gefundene Supraleitung verantwortlich sein.

Im magnetischen Kondo-Effekt koppeln die Leitungselektronen \"uber ihre Spindichte an das lokale magnetische Moment. Die dynamischen Korrelationen des lokalen Moments sind daher an die lokale dynamische Spinsuszeptibilit\"at gekoppelt, die in Fermi-Fl"ussigkeiten ein in der Frequenz lineares Verhalten besitzt. Das legt nahe, dass man unter geeigneten Umst\"anden das fermionische Bad durch ein sogenanntes Ohmsches Bad, wie es akustische Phononen darstellen, ersetzen kann. Tats\"achlich kann man das anisotrope Kondo-Modell auf das Ohmsche Spin-Bosonen-Modell abbilden, in dem ein Spin an ein Bad von akustischen Bosonen gekoppelt ist. M\"ogliche Realisierungen dieses Modells werden wir im n\"achsten Abschnitt kurz besprechen.

In \"ahnlicher Weise kann Kondo-Streuung in rein phononischen Systemen auftreten. Ein Beispiel sind sogenannte Clathratverbindungen, in denen in molekularen K\"afigen nur schwach gebundene Atom starke lokale Schwingungen (``rattling''-Moden) ausf\"uhren. Aufgrund der Symmetrie des K\"afigs sind zwei der Moden entartet und k\"onnen somit einen Pseudospin definieren, der nun \"uber die Ankopplung eines Bads akustischer Phononen Kondo-abgeschirt werden kann \cite{Ike19.1} (Abbildung \ref{Fig_other_Kondo}\,d).\\[2ex]

{\Large\bf Kondo-Effekt in k\"unstlichen Strukturen}\\

\begin{figure}[!h]
	\centering
	\includegraphics[width=0.95\linewidth]{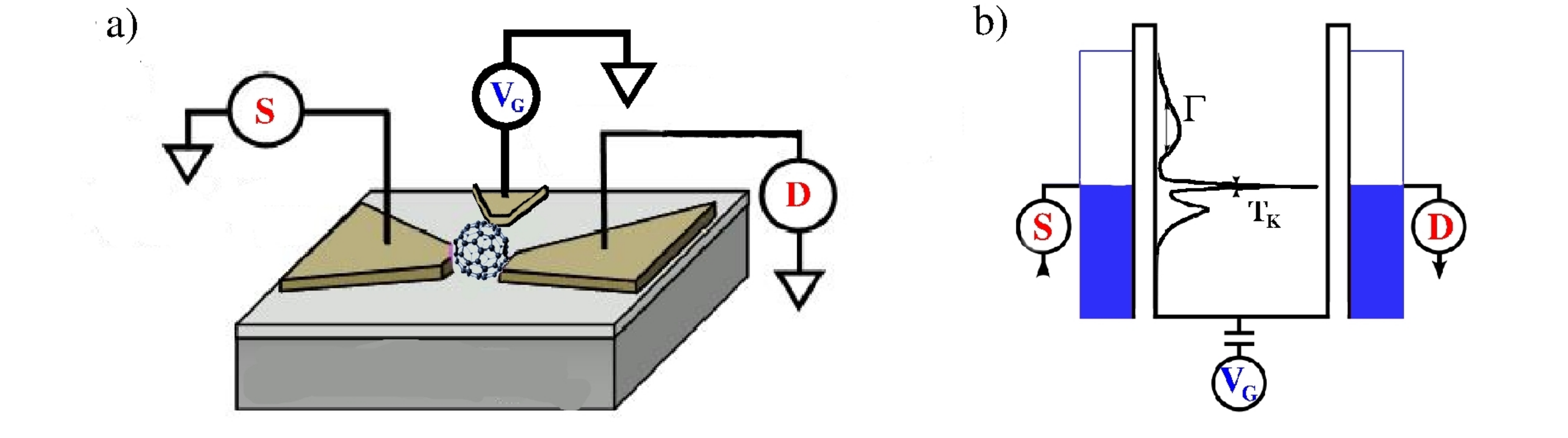}
	\caption{{\bf KONDO-EFFEKT IN NANOSTRUKTUREN\\
 	a) Schematische Darstellung eines Einelektronentransistors, bestehend aus einem Molek\"ul, das an zwei	Zuleitungen (S und D) gekoppelt ist. Eine Gatterelektrode ($V_{\rm G}$) erlaubt es, lokale Energieskalen und damit die resultierende Kondo-Temperatur 				durchzustimmen. b) Die resultierende lokale Zustandsdichte des Transistors zeigt die Kondo-Resonanz als scharfe Struktur in der N\"ahe der Fermi-Energie der Zuleitungen. Ihre Breite entspricht in etwa der Kondo-Temperatur $T_{\rm K}$} (nach \cite{2CKReview}).}
	\label{Kondo_nano}
\end{figure}

\noindent Der Kondo-Effekt tritt nicht nur in Quantenmaterialien auf, sondern auch in ``k\"unstlichen'' (nicht in der Natur existierenden) Strukturen. Zun\"achst theoretisch  vorhergesagt, wurde Kondo-Streuung experimentell unter anderem in Halbleiterheterostrukturen, atomaren Ketten, Nanor\"ohrchen und Molek\"ulkomplexen beobachtet. Die Grundlage von Kondo-Physik in solchen Strukturen ist in Abbildung \ref{Kondo_nano}\,a skizziert. Ein Quantentopf (z.B.\ in Form eines Molek\"uls) ist an Zuleitungen sowie an eine Gatterelektrode gekoppelt. Die Zuleitungen dienen hierbei als Reservoirs f\"ur Leitungselektronen (bzw.\ Teilchen-Loch-Anregungen), die den lokalen Spin, der sich im Quantentopf ausbildet, abschirmen.  Die f\"ur den Kondo-Effekt relevanten Energieskalen sind die Aufspaltung $\Delta$ der Energie-Niveaus im Quantentopf und die Ladungsenergie $E_C$, die n"otig ist, um die Zahl der Elektron im Quantentopf um ein Elektron zu ver\"andern. Die Aufspaltung $\Delta$ ist eine Konsequenz der endlichen Gr\"o{\ss}e des Topfes. Unter gewissen Umst\"anden, die sich mittels Gatterelektrode kontrollieren lassen, sind Ladungsfluktuationen mit den Zuleitungen unterdr\"uckt. In diesem Regime kann sich der Spin-Kondo-Effekt ausbilden.

Abbildung \ref{Kondo_nano}\,b zeigt die lokale Zustandsdichte des Quantentopfes im Kondo-Regime. Deutlich zu sehen ist eine Resonanz in der N\"ahe der Fermi-Energien der Zuleitungen. Dies ist die Kondo-Resonanz, deren charakteristische Breite durch $T_{\rm K}$ gegeben ist. Im Festk\"orper f\"uhrt der Kondo-Effekt zu einer Erh\"ohung des Widerstands, weil die Leitungselektronen an den magnetischen St\"orstellen streuen. In Nanostrukturen hingegen unterst\"utzt die durch den Kondo-Effekt erh\"ohte lokale Zustandsdichte den Transport von Ladungen durch den Quantentopf. Daher zeigt sich der Kondo-Effekt durch eine Erh\"ohung der Leitf\"ahigkeit, die weit unterhalb von $T_{\rm K}$ den maximal m\"oglichen Wert von $e^2/h$ pro Kanal (also insgesamt $2e^2/h$) annimmt. Dies ist der unit\"are Limes.

In Nanostrukturen l\"asst sich auch eine besondere Art des Ladungs-Kondo-Effekts realisieren und zwar dann, wenn die Gatterspannung gerade so gro{\ss} ist, dass die Ladungszust\"ande des Quantentopfs mit $n$ und $n+1$ Elektronen entartet sind. Dann entspricht das Tunneln von Elektronen in den Quantentopf und wieder hinaus der Spinaustauschstreuung im Spin-Kondo-Effekt und der Ladungszustand des Quantentopfs kann als Pseudospins beschrieben werden.

Das Ohmsche Spin-Boson-Modell (s.o.), das auf das anisotrope Kondo-Modell abgebildet werden kann, wird oft im Kontext der {\em circuit quantum electrodynamics} (C-QED) verwendet (es tritt hier als Verallgemeinerung der  Rabi- und Jaynes--Cummings-Modelle auf). Experimentell kann es mit Qubits, die an ein bosonische Bad mit linearer Dispersion ankoppeln, realisiert werden. Das Bad kann eine eindimensionale Anordnung von abstimmbaren Josephson-Kontakten sein oder auch die niederenergetischen Schallanregungen eines Bose-Einstein-Kondensats.

Vorschl"age, das Kondo- und das Kondo-Gitter-Modell in ultrakalten Gasen zu realisieren, existieren seit mehreren Jahren und mittlerweile ist es zumindest gelungen, eine durchstimmbare Spinaustauschwechselwirkung in einem System ultrakalter $^{173}$Yb-Atome zu realisieren \cite{Riegger.18}. Weiteren neuen Realisierungen von effektiven Kondomodellen sind offenbar
nur durch die Kreativit"at Grenzen gesetzt.\\[2ex]
		
\newpage

{\Large\bf Kondo mit Twist}\\
		
\begin{figure}[!h]
	\hspace{-7cm}\large{$j_y \sim \int d{\bf k} g({\bf k}, E_x)\Omega_z({\bf k}) E_x$}\\
	\centering
	\includegraphics[width=0.6\linewidth]{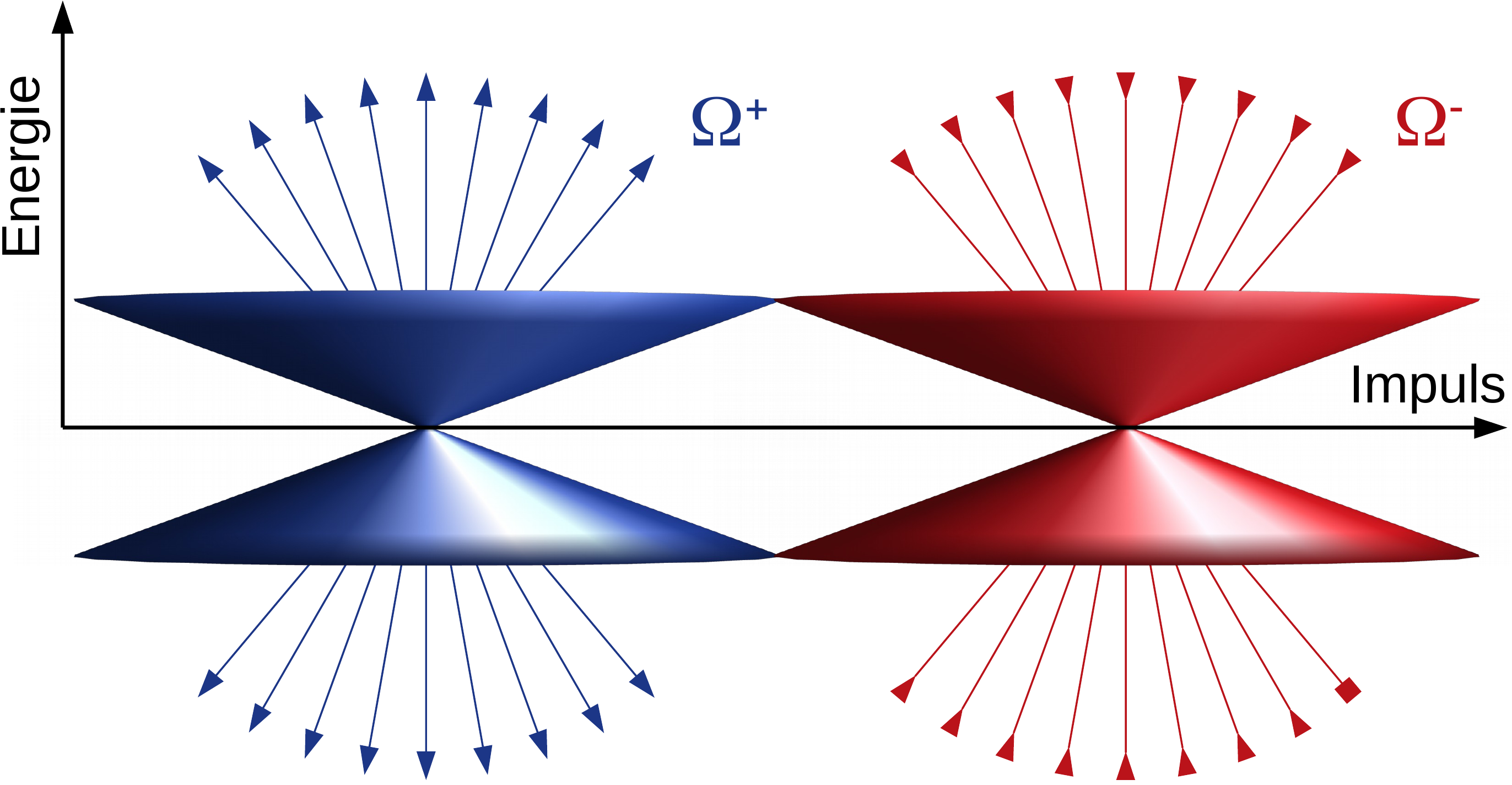}\hspace{0.6cm}
	\includegraphics[width=0.31\linewidth]{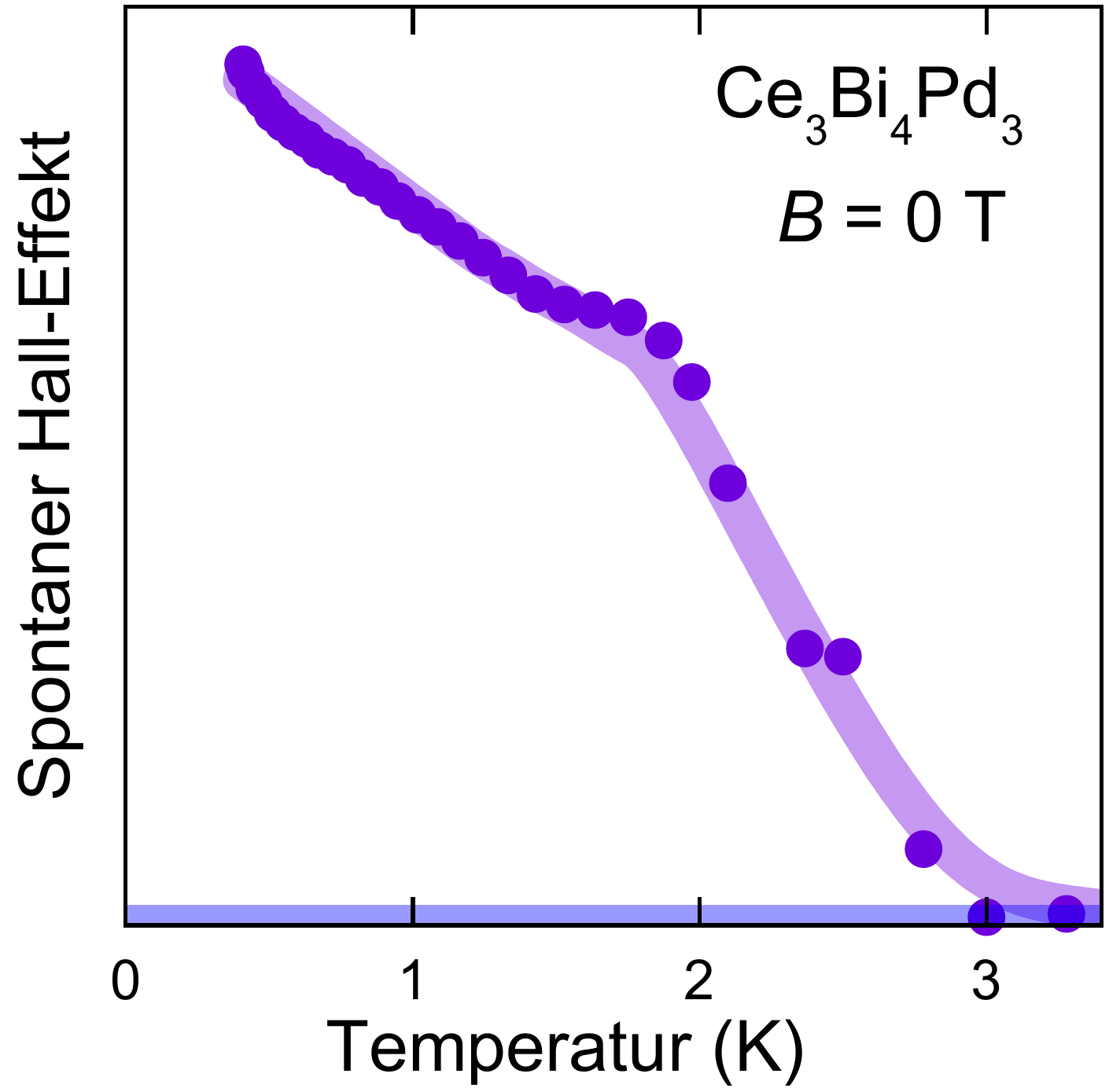}
	\caption{{\bf WEYL-KONDO-HALBMETALLE\\
	Der Energie--Impuls-Zusammenhang (Dispersion, linkes Teilbild) ist in der Umgebung der Weyl-Punkte linear (blauer und roter Doppelkegel). Dieses Verhalten wird mit der Dirac-Gleichung statt wie sonst f\"ur Elektronen im Festk\"orper \"ublich mit der Schr\"odinger-Gleichung beschrieben. In einem Weyl-Kondo-Halbmetall sind die Weyl-Kegel Teil der Kondo-Resonanz. Die Weyl-Punkte liegen daher in unmittelbarer N\"ahe zur Fermi-Energie und die Weyl-Kegel sind extrem flach. In einem externen elektrischen Feld ($E_x$) resultiert aus der Wirkung der Berry-Kr\"ummung ($\Omega_z$) auf die Nichtgleichgewichtsverteilung der Elektronen ($g$) im Impulsraum ($\bf k$) ein senkrechter Stromfluss ($j_y$) und damit ein (spontaner) Halleffekt. Dieser ist f\"ur das Weyl-Kondo-Halbmetall Ce$_3$Bi$_4$Pd$_3$ gezeigt (rechtes Teilbild). Die Hallspannung entsteht erst bei tiefen Termperaturen (hier unterhalb von 3\,K), wenn der (koh\"arente) Kondo-Gitter-Effekt voll ausgebildet ist} (nach \cite{Dzs21.1}).}
	\label{Weyl}
\end{figure}

\noindent Ein v\"ollig neues Kapitel in der Kondo-Physik wurde mit der Entdeckung aufgeschlagen, dass der Kondo-Effekt zu topologischen Zust\"anden mit ungeahnten Eigenschaften f\"uhren kann. Topologische Phasen werden anders als Phasen mit gebrochener Symmetrie (z.B. dem oben beschriebenen Antiferromagneten) nicht mit einem lokalen Ordnungsparameter beschrieben, sondern mittels sogenannter topologischer Invariante, die Phasen global charakterisieren. Ein besonders interessantes Beispiel einer topologischen Phase ist das Weyl-Halbmetall \cite{Nie18.1}. In einem solchen Material verhalten sich die Elektronen auch schon ohne den Kondo-Effekt sehr ungew\"ohnlich. So steht z.B.\ ihr Spin in einer ganz bestimmten Beziehung zu ihrem Impuls, und ihre Energie-Impuls-Beziehung (als Dispersion bezeichnet, Abb.\,\ref{Weyl}) ist in der Umgebung der Weyl-Punkte linear statt wie in einfachen Leitern \"ublich quadratisch \cite{Nie18.1}. Die Weyl-Punkte, bei denen sich das Leitungsband und das Valenzband treffen, sind durch die Symmetrie des Materials stabilisiert und k\"onnen nur dann verschwinden, wenn sich zwei zusammengeh\"orende Weyl-Punkte im Impulsraum treffen---ein Ph\"anomen, das der Annihilation eines Weyl-Teilchens mit einem Anti-Weyl-Teilchen in der Teilchenphysik entspricht (wo dies bisher allerdings nicht beobachtet werden konnte). Diese Weyl-Punkte wirken auf Leitungselektronen im Festk\"orper wie magnetische Monopole.
		
Wie k\"urzlich gezeigt wurde, kann der Kondo-Effekt dies ganz besonders deutlich zutage bringen. F\"ur die meisten Eigenschaften im Festk\"orper sind nur die Elektronen am Fermi-Niveau verantwortlich. Um den exotischen Effekt der Weyl-Punkte messbar zu machen, m\"ussen die Weyl-Punkte energetisch daher sehr nahe am Fermi-Niveau liegen, und genau dies kann der Kondo-Effekt bewerkstelligen. Die Weyl-Fermionen werden Teil der Kondo-Resonanz und liegen damit so wie die Kondo-Resonanz selbst in unmittelberer N\"ahe zum Fermi-Niveau. Flie{\ss}t in einem solchen als Weyl-Kondo-Halbmetall bezeichneten Material ein elektrischer Strom, so werden die Elektronen durch das fiktive Magnetfeld stark abgelenkt. Es entsteht ohne jegliches extern angelegte Magnetfeld eine Hallspannung (Abbildung \ref{Weyl}). Obwohl dieser ``spontane'' Halleffekt auch in Weyl-Halbmetallen ohne Kondo-Wechselwirkung existieren sollte, ist das Signal hier offenbar zu klein um nachgewiesen zu werden.

{\Large\bf Kondo ganz verschr\"ankt}\\

\noindent In der makroskopischen Welt ist der Zustand, den ein Objekt annimmt, wohl determiniert. Eine Katze zum Beispiel ist entweder tot oder sie lebt. Quantenmechanische Objekte hingegen k\"onnen in einer Superposition von mehreren Zust\"anden vorliegen (so wie im Gedankenexperiment die Schr\"odinger-Katze, die zugleich lebt und tot ist, solange wir den Schrank nicht \"offnen). Das damit verbundene Ph\"anomen der Quantenverschr\"ankung ist ein Hauptmerkmal von Quantensystemen. So stellt zum Beispiel die Verschr\"ankung von zwei Qubits das Herzst\"uck eines Quantencomputers dar.

Auch der Kondo-Effekt ist ein Verschr\"ankungseffekt. Zu einer besonders interessanten Erkenntnis kam hier k\"urzlich eine theoretische Studie. Genau dort, wo der Kondo-Effekt aufbricht, wird die Verschr\"ankung langreichweitig, n\"amlich an einem quantenkritischen Punkt mit Kondo-Aufbruch \cite{Wag18.1}. In allen anderen F\"allen, so die Theorie, sollte die Kondo-Verschr\"ankung nur eine begrenzte, relative kurze Reichweite haben. Experimente sollten dies freilich best\"atigen. Die Voraussetzungen, solche Messungen tats\"achlich durchf\"uhren zu k\"onnen, wurden in letzter Zeit geschaffen. Von der Verbindung YbRh$_2$Si$_2$, in der wie oben diskutiert der quantenkritische Punkt mit Kondo-Aufbruch besonders gut etabliert ist (s.o.), existieren inzwischen per Molekularstrahlepitaxie (MBE) hergestellte D\"unnfilme \cite{Pro20.1}. Dies erlaubt die Strukturierung des Materials, wie sie zum Beispiel zur Durchf\"uhrung von Verschr\"ankungsmessungen mittels sogenannter Bipartit-Fluktuationen \cite{Son12.1} n\"otig ist. Hier wird ein Material in zwei Bereiche unterteilt, und die Fluktuationen (z.B.\ des Spins oder der Ladung) werden als Funktion der L\"ange der Grenzlinie zwischen beiden Bereichen untersucht. Gelingen solche Messungen, so k\"onnte dies zur Entwicklung eines Verschr\"ankungs-Schalter verwendet werden: Mit nur minimaler \"Anderung des externen Kontrollparameters (z.B.\ des Magnetfelds im Falle von YbRh$_2$Si$_2$) k\"onnte die langreichweitige Verschr\"ankung ein und aus geschaltet werden.

Auch mit anderen Entwicklungen in Kondo-Systemen bewegt man sich in Richtung Quantenanwendungen. So sind z.B.\ einige der vielversprechendsten Spin-Triplett-Supraleiter Schwerfermionen-Systeme und stabile Vielteilchen-Energiel\"ucken sind auch in Kondo-Isolatoren zu finden. In beiden Materialklassen finden sind Kandidaten zur Realisierung der hei{\ss} begehrten Majorana-Fermionen -- mittels derer fehlertolerante Quantenrechner gebaut werden sollen. Zu ihrem Nachweis wie auch zur Realisierung von Majorana-{\em Devices} d\"urften MBE-Filme wieder eine zentrale Rolle spielen.\\[2ex]

{\Large\bf ZUSAMMENFASSUNG}\\

\noindent Der Kondo-Effekt ist ein Vielteilchen-Effekt, der die Eigenschaften ganz unterschiedlicher Materialklassen pr\"agt und somit zentrales Thema in der Festk\"orperphysik ist. Schon der urspr\"ungliche Spin-Kondo-Effekts ist nicht nur wegen des Kondo-abgeschirmten Zustands mit hohen effektiven Massen von Interesse, sondern auch wegen der Physik des Kondo-Aufbruchs -- dem quantenkritischen Ladungsdelokalisierungs\"ubergang mit {\em strange metal}-Verhalten und unkonventioneller Supraleitung. In Kombination mit starker Spin-Bahn-Kopplung treten neuartige topologischen Ph\"anomenen auf. In einer verallegemeinterten Betrachtung k\"onnen andere Zwei- oder sogar Mehr-Niveausysteme an die Stelle des lokalisierten Spins treten und sogar akustische Phononen die Rolle der Ladungstr\"ager \"ubernehmen. Kondo-Physik tritt auch in ``k\"unstlichen'' Systemen auf und an Implementierungen in kalten Atomsystemen wird gearbeitet. Neueste Entwicklungen deuten darauf hin, dass der Kondo-Effekt auch zur Entwicklung quantentechnologischer Systeme f\"uhren k\"onnte.\\[2ex]	


\phantom{.}
\vskip 20pt
\newpage

{\bf KURZBIO}\\[1.5ex]
\begin{minipage}{.2\textwidth}
	\includegraphics[width=3cm]{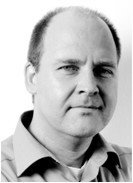}
\end{minipage}
\begin{minipage}{.75\textwidth}
{\bf Stefan Kirchner}\\
\Letter ~~Zhejiang Institute of Modern Physics \& Department of Physics\\ Zhejiang University\\ Hangzhou, 310027, China\\
    \url{www.correlated-matter.com}\\
\Email ~~stefan.kirchner@correlated-matter.com\\[1.2ex]
Stefan Kirchner ist Professor f"ur Physik an der Zhejiang Universit"at. Er studierte an der Universit"at W\"{u}rzburg und der State University of New York, USA und promovierte an der Technischen Universit"at Karlsruhe. Sein Arbeitsgebiet ist die theoretische Physik kondensierter Materie mit einem Schwerpunkt auf stark korrelierte Systeme innerhalb und fern des thermodynamischen Gleichgewichts.
\end{minipage}
\phantom{.}
\vskip 30pt
\begin{minipage}{.2\textwidth}
	\includegraphics[width=3cm]{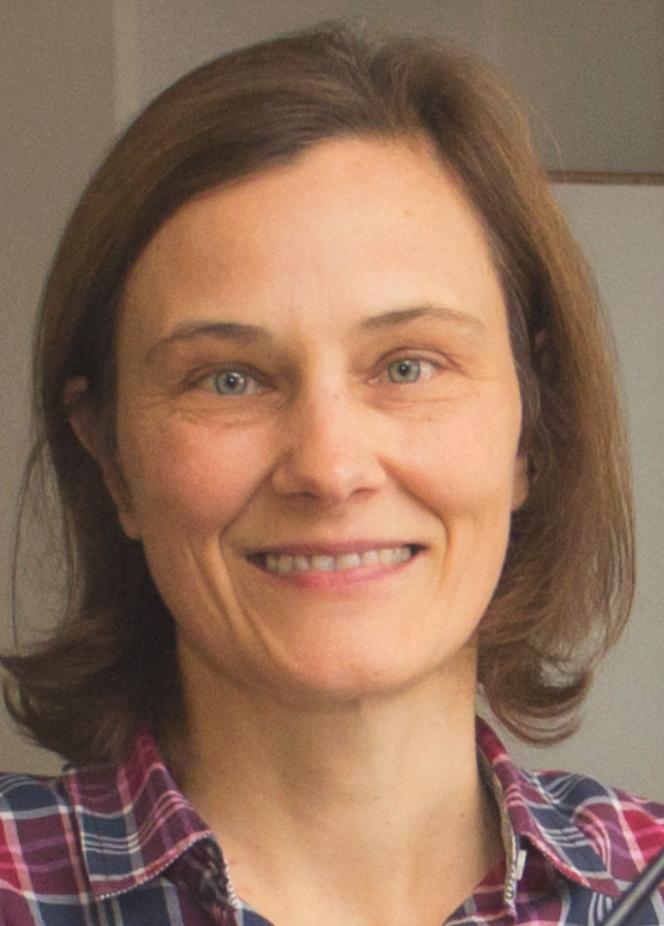}
\end{minipage}
\begin{minipage}{.75\textwidth}
{\bf Silke (B\"uhler-)Paschen}\\
\Letter ~~Institut f\"ur Festk\"orperphysik, Technische Universit\"at Wien\\ 1040 Wien, \"Osterreich\\
    \url{www.ifp.tuwien.ac.at/paschen/}\\
\Email ~~paschen@ifp.tuwien.ac.at\\[1.2ex]
Silke Paschen ist Physik-Professorin an der TU Wien. Sie studierte an der TU Graz und promovierte an der EPFL in Lausanne. Ihr Arbeitsgebiet ist die experimentelle Erforschung von stark korrelierten Elektronensystemen, wobei ein breites Spektrum an Methoden -- von der Synthese bis hin zu Messungen bei ultratiefen Temperaturen -- zum Einsatz kommt. 
\end{minipage}

\phantom{.}
\vskip 30pt

\begin{wrapfigure}{r}{17.5cm}  
\begin{minipage}[t]{0.97\linewidth} 
\begin{tcolorbox}[colback=gray!5,colframe=green!40!black,title=Kondo-Effekt und die Orthogonalit\"atskatastrophe]
Bei der Streuung eines Elektrons an einem punktf"ormigen Potential $V$
unterscheidet sich die einlaufende ($\Psi_{V=0}\sim\frac{\sin(kr)}{r}$) von der gestreuten sph"arischen Welle  ($\Psi_{V}\sim\frac{\sin(kr+\delta)}{r}$) nur durch die Streuphase   $\delta$. Es folgt, dass der Grundzustand eines Fermigases mit  lokalem St"orpotential ($|\mbox{FS}(V)\rangle $) orthogonal zum Grundzustand ohne das Str"orpotential ($|\mbox{FS}(V=0)\rangle $) ist. Konkret besagt die Andersonsche Orthogonalit"atskatastrophe, dass $\langle \mbox{FS}(V)|\mbox{FS}(V=0)\rangle \sim 1/N^{\delta^2/\pi^2}$, was  f"ur hohe Teilchenzahlen $N$ verschwindet. Dies ist eine Konsequenz der Anh\"aufung von Teilchen-Loch-Paaren mit verschwindender Anregungsenergie.
	
Der Kondo-Effekt tritt bei der Streuung von Elektronen an St\"orstellen mit internen, quantenmechanischen Freiheitsgraden auf. Bei jedem solchen Streuprozess induziert die \"Anderung im Streupotential eine Orthogonalit\"atskatastrophe und logarithmische Singularit\"aten. Die Kondo-Austauschwechselwirkung zwischen dem Moment $\mathbf{S}$ am Ort $\vec{r}=0$ und der Leitungselektronenspindichte  $\mathbf{s}(\vec{r}=0)$ hat die Form $J_{\rm K} \mathbf{S}\cdot \mathbf{s}(\vec{r}=0)$. Renormierungsgruppenrechnungen zeigen, dass
\begin{equation}
\label{eq:PMS1}
\frac{dJ}{dD}= -N_0 \frac{J^2}{D} \; ,
\end{equation}
gilt, wobei $N_0$ die elektronische Zustandsdichte an der Fermi-Energie ist. Diese Gleichung besagt, dass eine \"Anderung der Leitungselektronenbandbreite $D\rightarrow D+dD$ einer entsprechenden \"Anderung der Austauschwechselwikung $J$ bedarf, sodass die Streumatrix des Problems f\"ur kleine Energien ($ \ll D+dD$) invariant ist. F\"ur ferromagnetische Kopplungen ($J_{\rm K}<0$) verschwindet $J_{\rm K}$ ($J_{\rm K} \rightarrow 0$), w\"ahrend es f\"ur antiferromagnetische Kopplungen ($J_{\rm K}<0$) diverigert ($J_{\rm K} \rightarrow \infty$ f\"ur $D\rightarrow 0$). $J \rightarrow \infty$  signalisiert die Bildung des Singulett-Zustandes und damit den Kondo-Effekt.
	
Dies ist ein Beispiel f\"ur die sogenannte asymptotische Freiheit, die aus der Quantenchromodynamik bekannt ist: f\"ur hohe Temperaturen kann die Kopplung des Spinfreiheitsgrads an die Leitungselektronen vernachl\"assigt werden; der Spin ist frei, was sich in einer Curie-artigen Spinsuszeptibilit\"at wiederspiegelt. F\"ur Energien oder Temperaturen weit unterhalb der Kondo-Temperatur ist der Spinfreiheitsgrad  durch die Singulett-Bildung eingeschr\"ankt ({\itshape confined}). Numerische Renormierungsgruppenrechnungen zeigen, dass die Anregungen \"uber dem Singulett-Grundzustand fermionischer Natur sind. 
\end{tcolorbox} 
\end{minipage}  
\end{wrapfigure}
\phantom{.}
\newpage

\begin{wrapfigure}{r}{17.5cm}  
\begin{minipage}[t]{0.97\linewidth} 
\begin{tcolorbox}[colback=gray!5,colframe=green!40!black,title=Quantenkritikalit"at und dynamisches Skalenverhalten]
In der N"ahe von kontinuierlichen Phasen"uberg"angen zeigen physikalische Gr"o"sen Potenzverhalten\cite{Voj01.1}. 
Klassische Phasen"uberg"ange finden bei nichtverschwindenden Temperaturen ($T_{\rm c} > 0$) statt und werden durch einen Ordnungsparameter und dessen Fluktuationen beschrieben. Im Gegensatz dazu k"onnen nicht alle Quantenphasen"uberg"ange, die bei $T=0$ stattfinden, allein durch Ordnungsparameterfluktuationen beschrieben werden. F"ur diejenigen, f"ur die dies m"oglich ist, "ahnelt die theoretische Beschreibung jener der klassischen Phasen"uberg"ange. Hier k"onnen demnach Aussagen getroffen werden, unter welchen Umst"anden und in welchen Gr"o"sen Energie-"uber-Temperatur-Skalierung auftritt. 

Bei klassischen Phasen"uberg"angen verh\"alt sich die spezifische W"arme $C$ typischerweise wie $C\sim |t|^{-\alpha}$ und die Korrelationsl"ange $\xi$, ein Ma"s f"ur die Reichweite der Fluktuationen, divergiert am kritischen Punkt gem\"a{\ss} $\xi\sim |t|^{-\nu}$ . Der Parameter $t=(T-T_{\rm c})/T_{\rm c}$ ist ein Ma"s f"ur den Abstand zum Phasen"ubergang. Die kritischen Exponenten  ($\alpha,\nu,\ldots$) h"angen  im allgemeinen nicht von Details ab.

Innerhalb der N"aherung der Landau-Theorie, die einen nichtwechselwirkenden kritischen Punkt beschreibt, findet man $\alpha=0$, $\nu=1/2$. Allerdings weiss man, dass diese kritischen Exponenten nur in Raumdimensionen $d > 4$ korrekt sind, nicht aber f"ur zwei- und dreidimensionale magnetische Systeme. Hyperskalierungsbeziehungen (z.B.\ $d\nu=2-\alpha$) werden im allgemeinen von den Landau-Exponenten verletzt, au{\ss}er f"ur den Fall der oberen kritischen Dimension $d=4$. Hier fallen der nichtwechselwirkende (Gausssche) Fixpunkt und der wechselwirkende kritische Fixpunkt, der unterhalb von $d=4$ die kritische Physik bestimmt, zusammen.

Auf magnetische {\em Quanten}phasen"uberg"ange (bei $T=0$) angewandt implizieren diese Hyperskalierungsbeziehungen, dass die dynamische magnetische Suszeptibilit"at als Funktion der Energie ($E$) und der Temperatur die gleichen kritischen Exponenten zeigen muss, was "aquivalent zum $E/T$-Skalenverhalten ist. Allerdings muss beim \"Ubertragenen dieser Ordnungsparametertheorie auf $T=0$ die Raumdimension $d$ durch eine effektive Raumdimenion $D=d+z$ ersetzt werden, wobei $z$ der sogenannte dymanische Exponent ist. F"ur magnetische Quantenphasen"uberg"ange in Schwerfermionen-Systemen findet man f"ur antiferromagnetische "Uberg"ange $z=2$. Daher sollte f"ur die effektive Dimension $D=d+2>4$ erf"ullt sein und man erwartet Landau-Exponenten, keine Hyperskalierung und eine damit einhergehende Verletzung des dynamischen Skalenverhaltens in $E/T$.  Wird dennoch $E/T$-Skalenverhalten experimentell beobachtet, deutet dies daher darauf hin, dass kritische Prozesse jenseits von Ordnungsparamterfluktuationen die Physik bestimmen.
\end{tcolorbox} 
\end{minipage}  
\end{wrapfigure}

\end{document}